\documentclass[superscriptaddress,twocolumn,english,floatfix,aps,prd,amsmath,amssymb,longbibliography,10pt]{revtex4-2}
\usepackage[T1]{fontenc}
\usepackage[latin9]{inputenc}
\usepackage{color}
\usepackage{times}
\usepackage{babel}
\usepackage{amsmath}
\usepackage{amssymb}
\usepackage{graphicx}
\usepackage[colorlinks=true]{hyperref}
\hypersetup{citecolor=blue,linkcolor=blue,urlcolor=blue}

\begin{document}

\title{Dynamical Casimir effect enhanced by decreasing the mirror reflectivity}

\author{Andreson L. C. Rego}
\email{andresonlcr@gmail.com}
\affiliation{Instituto de Aplicação Fernando Rodrigues da Silveira, Universidade
do Estado do Rio de Janeiro, 20261-005, Rio de Janeiro, Brazil}

\author{Alessandra N. Braga}
\email{alessandrabg@ufpa.br}
\affiliation{Instituto de Estudos Costeiros, Universidade Federal do Pará, 68600-000,
Bragança, Brazil}

\author{Jeferson Danilo L. Silva}
\email{jdanilo@ufpa.br}
\affiliation{Campus Salinópolis, Universidade Federal do Pará, 68721-000, Salinópolis,
Brazil}

\author{Danilo T. Alves}
\email{danilo@ufpa.br}
\affiliation{Faculdade de Física, Universidade Federal do Pará, 66075-110, Belém,
Brazil}
\affiliation{Centro de F\'{i}sica, Universidade do Minho, P-4710-057, Braga, Portugal}
%\date{\today}

\begin{abstract}
In the present paper, we show that a partially reflecting static mirror with time-dependent
properties can produce, via dynamical Casimir effect in the context of a massless scalar field in $1+1$ dimensions, a larger number of particles than a perfectly reflecting one.
As particular limits, our results recover those found in the literature
for a perfect static mirror imposing a generalized or an usual time-dependent Robin boundary condition. 
\end{abstract}
\maketitle

%%%%%%%%%%%%%%%%%%%%%%%%%%%%%%%%%%%%%%%%%%%%%%%%%%%%%%%%%%%%%%%%%%%%
\section{Introduction}

The creation of real particles by excitation of the quantum vacuum by a moving mirror
was predicted by Moore \cite{Moore-1970}, and investigated
in other pioneering works in the 1970s \cite{DeWitt-PhysRep-1975,Fulling-Davies-PRSA-1976,Davies-Fulling-PRSA-1977,Candelas-PRSA-1977}.
Nowadays, this effect is commonly called the dynamical Casimir effect (DCE),
a name adopted by Yablonovitch \cite{Yablonovitch-PRL-1989} and Schwinger \cite{Schwinger-PNAS-1992}, motivated by a certain similarity with another quantum vacuum
effect involving mirrors, the so-called Casimir effect \cite{Casimir-1948}.
On the DCE there are some excellent reviews \cite{Dodonov-JPCS-2009, *Dodonov-PhysScr-2010, *Dalvit-CasimirPhysics-2011,Dodonov-Phys-2020}.
Moreover, a moving mirror is a particular way to excite the quantum vacuum. 
An alternative way was proposed by Yablonovitch \cite{Yablonovitch-PRL-1989} and Lozovik \textit{et al.} \cite{Lozovik-PZhETF-1995}, consisting in exciting the vacuum by means of time-varying properties of material media, which can simulate a moving mirror.
Several scientists have focused on the detection of photon creation
from mechanically moving mirrors \cite{Kim-PRL-2006,*Brownell-JPA-2008,*Macri-PRX-2018,*Motazedifard-2018,*Sanz-Quantum-2018,
*DiStefano-PRL-2019,*Qin-PRA-2019,*Butera-PRA-2019}, but the observation remains a challenge \cite{Dodonov-Phys-2020}.
Other ones have focused on simulating a moving mirror through a motionless 
medium whose internal properties vary in time 
\cite{Braggio-EPL-2005, *Braggio-JPA-2008, *Braggio-JPCS-2009, 
*Dezael-Lambrecht-EPL-2010, Wilson-Nature-2011,Kawakubo-Yamamoto-PRA-2011, 
*Faccio-Carusotto-EPL-2011,*Johansson-et-al-PRL-2013,*Lahteenmaki-PNAS-2013, *Motazedifard-2015,*Dodonov-JPA-2015,*Ugalde-PRA-2016,*Braggio-JOpt-2018,*Dodonov-IOP-CS-2019,*Vezzoli-CommPhys-2019,*Schneider-et-al-PRL-2020,Johansson-PRL-2009,*Johansson-et-al-PRA-2010}, with the first observation of photon creation from vacuum reported 
by Wilson \textit{et al.}, in Ref. \cite{Wilson-Nature-2011}.

One-dimensional models have had an important role in the investigation 
of the DCE.
It was adopted by Moore \cite{Moore-1970}, DeWitt \cite{DeWitt-PhysRep-1975},
Fulling and Davies \cite{Fulling-Davies-PRSA-1976}, and also
in many other works as, for instance, in Refs. \cite{Dodonov-Klimov-PRA-1996,*Alves-Farina-MaiaNeto-JPA-2003,*Alves-Farina-Granhen-PRA-2006,
*Alves-Granhen-PRA-2008, *Alves-Granhen-Lima-PRD-2008, *Alves-Granhen-Silva-Lima-PLA-2010, *Alves-Granhen-Silva-Lima-PRD-2010,*Alves-Granhen-CPC-2014,*Good-PRD-2016,*Good-PRD-2020,*Good-Orlando-ArXiv-2021,
Silva-Farina-PRD-2011,Silva-Braga-Rego-Alves-PRD-2015,
Silva-Braga-Alves-PRD-2016,Silva-Braga-Rego-Alves-PRD-2020}.
In $(1+1)$D, the simulation of a motionless mirror with internal properties varying in time
was proposed by Silva and Farina \cite{Silva-Farina-PRD-2011}, who considered
the quantum vacuum field submitted to a time-dependent Robin boundary condition on a static mirror. 
This model is deeply connected with the one underlying
the first experimental observation of photon creation, reported in Ref. \cite{Wilson-Nature-2011}.
The Robin boundary condition interpolates the well-known Dirichlet and Neumann ones \cite{Mintz-Farina-MaiaNeto-Rodrigues-JPA-2006-I,*Mintz-Farina-MaiaNeto-Rodrigues-JPA-2006-II}.
A generalized Robin boundary condition is one that includes a term
of second-order time derivative of the field, and has been
also considered in the investigation of the DCE 
\cite{Fosco-PRD-2013, Rego-Silva-Alves-Farina-PRD-2014}. 
One-dimensional models have also had relevance in the investigation of the DCE with semi-transparent mirrors.
Since real mirrors
do not behave as perfectly reflecting at all \cite{Moore-1970}, 
the DCE with partially reflecting mirrors has been investigated by 
several authors (see, for instance, \cite{Jaekel-Reynaud-Quant-Opt-1992,*Barton-Eberlein-AnnPhys-1993,*Lambrecht-Jaekel-Reynaud-PRL-1996,*Lambrecht-Jaekel-Reynaud-EPJD-1998,*Obadia-Parentani-PRD-2001,*Haro-Elizalde-PRL-2006,Barton-Calogeracos-AnnPhys-I-1995,Nicolaevici-CQG-2001,Nicolaevici-PRD-2009,Fosco-Giraldo-Mazzitelli-PRD-2017}).
Dirac $\delta$ potentials for modeling partially reflecting moving mirrors
were considered, for instance,
in Refs. \cite{Barton-Calogeracos-AnnPhys-I-1995,Nicolaevici-CQG-2001,Dalvit-MaiaNeto-PRL-2000,
Nicolaevici-PRD-2009},
and also in the investigation of the static Casimir effect \cite{Castaneda-Guilarte-PRD-2013}.
In the limit of a perfectly reflecting mirror, the $\delta$ model leads to the situation of a Dirichlet boundary condition.
The use of $\delta-\delta^{\prime}$ potentials ($\delta^{\prime}$ is the derivative of the Dirac $\delta$) has also been considered as, for example, in Refs. \cite{Castaneda-Guilarte-PRD-2015,Silva-Braga-Alves-PRD-2016,Braga-Silva-Alves-PRD-2016,Silva-Braga-Rego-Alves-PRD-2020}.
In the limit of a perfectly reflecting mirror, the 
$\delta-\delta^{\prime}$ model leads to a situation where 
the field obeys the Robin boundary condition on one side, 
and the Dirichlet condition on the other side of the mirror \cite{Silva-Braga-Alves-PRD-2016}.

One of the goals of the present paper is to show that transparency can enhance the number of particles created via DCE, when compared to the limiting case of a perfect mirror.
This counterintuitive effect was shown in Ref. \cite{Silva-Braga-Alves-PRD-2016}
(and highlighted by Dodonov in the review in Ref. \cite{Dodonov-Phys-2020}), 
in the context of a partially reflecting moving mirror, simulated by a $\delta-\delta^{\prime}$ potential.
In the present paper, we show that even a static partially reflecting mirror, with time-dependent
properties, can produce a larger number of particles than a perfectly reflecting one.
Specifically, we investigate this in the context of a massless scalar field in $(1+1)$D,
with a mirror described by a time-dependent generalized $\delta-\delta^{\prime}$ model.
As particular limits, our results recover those found in the literature for a perfect
static mirror imposing a time-dependent Robin \cite{Silva-Farina-PRD-2011}, or
a generalized time-dependent Robin boundary condition \cite{Rego-Silva-Alves-Farina-PRD-2014}. 

This paper is organized as follows. 
In Sec. \ref{sec:model}, we present
the Lagrangian density of the model, and obtain the corresponding scattering coefficients. 
In Sec. \ref{sec:spectrum}, the spectrum and total rate of created particles are obtained. 
In Sec. \ref{sec:application}, we apply our formulas to
a typical oscillatory behavior considered in investigating the DCE.
In Sec. \ref{sec:final-remarks},
we make a brief summary of our results.

%%%%%%%%%%%%%%%%%%%%%%%%%%%%%%%%%%%%%%%%%%%%%%%%%
\section{\label{sec:model}The Model}

We consider a massless scalar field in $1+1$ dimensions in the presence
of partially reflecting static mirror with time-dependent material
properties. The mirror is simulated by a $\delta-\delta^{\prime}$
potential at $x=0$ coupled to the field, and the material properties
of the mirror are represented by the coupling parameters. The $\delta$
term is coupled to the field by a time-dependent parameter, $\mu(t)$,
and the $\delta^{\prime}$ one by a time-independent parameter, $\lambda_{0}$.
Moreover, it is included a modification in the kinetic term of the
Lagrangian density at $x=0$ (where the mirror is located), namely
\begin{eqnarray}
\mathcal{L} & = & \frac{1}{2}\left[1+2\chi_{0}\delta(x)\right](\partial_{t}\phi)^{2}-\frac{1}{2}(\partial_{x}\phi)^{2}\nonumber \\
&  & -[\mu(t)\delta(x)+\lambda_{0}\delta^{\prime}(x)]\phi^{2}(t,x),
\label{eq:model}
\end{eqnarray}
where $\chi_{0}$ is a constant parameter. The modified kinetic term
of Eq. (\ref{eq:model}) originates a second-order time-derivative
that appears in the generalized Robin boundary condition (BC) \cite{Fosco-PRD-2013,Rego-Silva-Alves-Farina-PRD-2014}.
The model described by Eq. \eqref{eq:model} generalizes that 
of a perfectly reflecting time-dependent Robin boundary condition, found in Ref. \cite{Silva-Farina-PRD-2011}, 
and that of a perfectly reflecting mirror imposing a generalized time-dependent Robin BC to the field, found in Ref. \cite{Rego-Silva-Alves-Farina-PRD-2014}.
It also generalizes the semi-transparent time-dependent model 
considered  Ref. \cite{Silva-Braga-Rego-Alves-PRD-2020}.
At the end of this section, we clarify our motivations
for choosing $\chi_0$ and $\lambda_0$ constant in time, whereas
$\mu$ is made time-dependent, and also connect this model with some physical situations. 

The field equation for this model is given by
\begin{eqnarray}
[1+2\chi_{0}\delta(x)]\partial_{t}^{2}\phi(t,x)-\partial_{x}^{2}\phi(t,x)\nonumber \\
+2[\mu(t)\delta(x)+\lambda_{0}\delta^{\prime}(x)]\phi(t,x) & = & 0,
\label{eq:field-equation}
\end{eqnarray}
which becomes the massless Klein-Gordon equation
\begin{equation}
\partial_{x}^{2}\phi(t,x)-\partial_{t}^{2}\phi(t,x)=0,\text{ for }x\neq0.
\label{eq:KG-equation}
\end{equation}
A particular case of this model, with $\lambda_{0}=0$, was considered
in Ref. \cite{Fosco-PRD-2013}.

It is convenient to rewrite the field as
\begin{equation}
\phi(t,x)=\Theta(x)\phi_{+}(t,x)+\Theta(-x)\phi_{-}(t,x),
\label{phi-00}
\end{equation}
where $\Theta(x)$ is the Heaviside step function, $\phi_{+}$ and
$\phi_{-}$ are 
\begin{equation}
\phi_{+}(t,x)=\varphi_{\text{out}}(t-x)+\psi_{\text{in}}(t+x),
\end{equation}
\begin{equation}
\phi_{-}(t,x)=\varphi_{\text{in}}(t-x)+\psi_{\text{out}}(t+x),
\end{equation}
and the labels ``out'' and ``in'' indicate, respectively, the outgoing and incoming
fields with respect to the mirror. Taking the Fourier
transform, we obtain
\begin{equation}
\phi_{+}(t,x)=\int\frac{\mathrm{d}\omega}{2\pi}\left[\tilde{\varphi}_{\text{out}}(\omega)\text{e}^{i\omega x}+\tilde{\psi}_{\text{in}}(\omega)\text{e}^{-i\omega x}\right]\text{e}^{-i\omega t},
\label{eq:A08}
\end{equation}
\begin{equation}
\phi_{-}(t,x)=\int\frac{\mathrm{d}\omega}{2\pi}\left[\tilde{\varphi}_{\text{in}}(\omega)\text{e}^{i\omega x}+\tilde{\psi}_{\text{out}}(\omega)\text{e}^{-i\omega x}\right]\text{e}^{-i\omega t}.
\label{eq:A09}
\end{equation}
After two successive integrations of Eq. (\ref{eq:field-equation})
across $x=0$, we obtain the following matching conditions
\begin{equation}
\tilde{\varphi}_{\text{out}}(\omega)+\tilde{\psi}_{\text{in}}(\omega)=+\frac{1+\lambda_{0}}{1-\lambda_{0}}[\tilde{\varphi}_{\text{in}}(\omega)+\tilde{\psi}_{\text{out}}(\omega)],
\label{eq:MC1}
\end{equation}
\begin{eqnarray}
\tilde{\varphi}_{\text{out}}(\omega)-\tilde{\psi}_{\text{in}}(\omega) & = & \frac{1-\lambda_{0}}{1+\lambda_{0}}\left[\tilde{\varphi}_{\text{in}}(\omega)-\tilde{\psi}_{\text{out}}(\omega)\right]\nonumber \\
&  & +\frac{2i\chi_{0}\omega}{1-\lambda_{0}^{2}}\left[\tilde{\varphi}_{\text{in}}(\omega)+\tilde{\psi}_{\text{out}}(\omega)\right]\nonumber \\
&  & -\frac{2i}{\omega(1-\lambda_{0}^{2})}\int\frac{\mathrm{d}\omega^{\prime}}{2\pi}\tilde{\mu}(\omega-\omega^{\prime})\nonumber \\
&  & \times\left[\tilde{\varphi}_{\text{in}}(\omega^{\prime})+\tilde{\psi}_{\text{out}}(\omega^{\prime})\right],
\label{eq:MC2}
\end{eqnarray}
where $\tilde{\mu}(\omega)$ is the Fourier transform of $\mu(t)$.

We shall consider
\begin{equation}
\mu(t)=\mu_{0}[1+\epsilon f(t)],
\label{eq:mu-of-t}
\end{equation}
where $\mu_{0}$ is a constant parameter, $f(t)$ is an arbitrary
limited function with $|f(t)|\le1$ and $\epsilon\ll1$. Moreover,
the outgoing and incoming fields are grouped in column matrices:
\begin{equation}
\Phi_{\text{out}}(\omega)=\left(\begin{array}{c}
\tilde{\varphi}_{\text{out}}(\omega)\\
\tilde{\psi}_{\text{out}}(\omega)
\end{array}\right),\quad\Phi_{\text{in}}(\omega)=\left(\begin{array}{c}
\tilde{\varphi}_{\text{in}}(\omega)\\
\tilde{\psi}_{\text{in}}(\omega)
\end{array}\right).
\end{equation}
Manipulating Eqs. (\ref{eq:MC1}) and (\ref{eq:MC2}), considering
Eq. (\ref{eq:mu-of-t}) and neglecting the terms $\mathcal{O}(\epsilon^{2})$,
the outgoing fields can be rewriten in term of the incoming ones,
namely
\begin{equation}
\Phi_{\text{out}}(\omega)=S(\omega)\Phi_{\text{in}}(\omega)+\int\frac{\mathrm{d}\omega^{\prime}}{2\pi}\mathcal{S}(\omega,\omega^{\prime})\Phi_{\text{in}}(\omega^{\prime}),
\label{eq:out-in}
\end{equation}
where $S(\omega)$ is the scattering matrix for the case $\mu(t)\rightarrow\mu_{0}$,
and $\mathcal{S}(\omega,\omega^{\prime})$ is the correction to the
scattering matrix due to the time-dependence of $\mu(t)$. Explicitly,
\begin{equation}
S(\omega)=\left(\begin{array}{cc}
s_{+}(\omega) & r_{+}(\omega)\\
r_{-}(\omega) & s_{-}(\omega)
\end{array}\right),
\label{eq:Scattering-matrix}
\end{equation}
where
\begin{equation}
s_{\pm}(\omega)=\frac{\omega(1-\lambda_{0}^{2})}{i\mu_{0}-i\chi_{0}\omega^{2}+\omega(1+\lambda_{0}^{2})},
\end{equation}
\begin{equation}
r_{\pm}(\omega)=-\frac{i\mu_{0}-i\chi_{0}\omega^{2}\mp2\omega\lambda_{0}}{i\mu_{0}-i\chi_{0}\omega^{2}+\omega(1+\lambda_{0}^{2})},
\label{r-pm}
\end{equation}
are the transmission and reflection coefficients, respectively. The
term $\mathcal{S}(\omega,\omega^{\prime})$ is given by 
\begin{equation}
\mathcal{S}(\omega,\omega^{\prime})=-\frac{i\epsilon\mu_{0}\tilde{f}(\omega-\omega^{\prime})\left[J_{2}+S(\omega^{\prime})\right]}{i\mu_{0}-i\chi_{0}\omega^{2}+\omega(1+\lambda_{0}^{2})},
\label{eq:S-full}
\end{equation}
where $\tilde{f}(\omega)$ is the Fourier transform of $f(t)$ and
$J_{2}$ is the $2\times2$ backward identity matrix. The scattering
matrix (\ref{eq:Scattering-matrix}) must be real in the temporal
domain, unitary and analytic for $\mathrm{Im}\,\omega>0$ \cite{Jaekel-Reynaud-Quant-Opt-1992,Lambrecht-Jaekel-Reynaud-PRL-1996},
which is guaranteed if $\mu_{0}$ and $\chi_{0}$ are non-negative.
Particularly, the limits $\mu_{0}\rightarrow\infty$ or $\chi_{0}\rightarrow\infty$
lead to the simpler case of a perfect mirror {[}$s_{\pm}(\omega)\rightarrow0${]}
imposing the Dirichlet BC to the field in both sides of the mirror,
namely
\begin{equation}
\phi(t,0^{+})=\phi(t,0^{-})=0,
\label{eq:Dirichlet-L-R}
\end{equation}
or, equivalently, $r_{\pm}(\omega)\rightarrow-1$, which leads to
$\mathcal{S}(\omega,\omega^{\prime})\rightarrow0$. On the other
hand, the limit $\lambda_{0}\rightarrow-1$ (or $\lambda_{0}\rightarrow1$)
also leads to a perfect mirror, but in this case $\mathcal{S}(\omega,\omega^{\prime})\neq0$,
and it leads to the BCs
\begin{equation}
\phi(t,0^{+})=0,
\label{eq:Dir}
\end{equation}
\begin{equation}
\mu(t)\phi(t,0^{-})+2\partial_{x}\phi(t,0^{-})+\chi_{0}\partial_{t}^{2}\phi(t,0^{-})=0.
\label{eq:time-dependent-generalized-Robin-BC}
\end{equation}
identified respectively as the Dirichlet BC (\ref{eq:Dir}) and the
generalized Robin BC with a time-dependent Robin parameter (\ref{eq:time-dependent-generalized-Robin-BC}).
Therefore, as we shall see, the results for the spectra of created
particles recover, in the appropriate limits, those
found in the literature \cite{Silva-Farina-PRD-2011, Rego-Silva-Alves-Farina-PRD-2014}.

Before continuing, let us make a brief comment about the nomenclature
we are using for the BC given in Eq. \eqref{eq:time-dependent-generalized-Robin-BC}.
Considering the particular case where $\mu(t)=0$ and $\chi_0=0$ in Eq. \eqref{eq:time-dependent-generalized-Robin-BC}, one has the Neumann BC, $\partial_{x}\phi(t,0^{-})=0$.
When we consider $\mu(t)=\mu_0>0$ and $\chi_0=0$, one has
\begin{equation}
	\mu_0\phi(t,0^{-})+2\partial_{x}\phi(t,0^{-})=0,
	\label{eq:robin-bc}
\end{equation}
which is a particular case of that usually called Robin's BC,
although G. Robin seems to have never used this BC (a very interesting
discussion on this subject is found in Ref. \cite{Gustafson-Math-Int-1998}).
For $\mu(t)>0$ and $\chi_0=0$, Eq. \eqref{eq:time-dependent-generalized-Robin-BC} gives
\begin{equation}
	\mu(t)\phi(t,0^{-})+2\partial_{x}\phi(t,0^{-})=0,
	\label{eq:time-dependent-robin-bc}
\end{equation}
which is a particular case of that called in Ref. \cite{Silva-Farina-PRD-2011} a \textit{time-dependent} Robin BC.
Following the nomenclatures adopted in the literature, the full BC given in Eq. \eqref{eq:time-dependent-generalized-Robin-BC} was called in Ref. \cite{Fosco-PRD-2013} as a \textit{generalized} Robin BC.

Concluding this section, we discuss our motivations for choosing $\chi_0$ and $\lambda_0$ constant in time, whereas $\mu$ is time dependent, and also
connect the model investigated here with physical situations. 
The BC in Eq. \eqref{eq:time-dependent-generalized-Robin-BC} is
related to the first observation of photon creation from vacuum,
which involved a superconducting coplanar waveguide, terminated at a
SQUID (superconducting quantum interference device) \cite{Wilson-Nature-2011}. 
In this case, the time-dependent parameter $\mu$ is 
related to a time-varying effective Josephson energy, whereas
$\chi_0$ is related to the constant capacitance and inductance per unit length
of the superconducting coplanar waveguide \cite{Johansson-PRL-2009, Johansson-et-al-PRA-2010}.
This motivated our choice to investigate a model where $\chi_0$ is a constant,
and $\mu$ is a time-dependent function. 
In addition, we chose the parameter $\lambda_0$ as a constant
whose value controls the reflectivity of the object in such a way that
when $\lambda_0=\pm 1$ one recovers a perfectly reflecting object, independently
of the values assumed by $\mu(t)$ and $\chi_0$ [as shown in Eq. \eqref{r-pm}].
Lastly, particular cases of the BC Eq. \eqref{eq:time-dependent-generalized-Robin-BC} are given in Eqs. \eqref{eq:robin-bc} and \eqref{eq:time-dependent-robin-bc}, and are also connected
with another physical situation related to the DCE.
The Robin BC in Eq. \eqref{eq:robin-bc} can be used to describe
a plasma model for a real metal,
with the parameter $\mu_0$ related with the plasma frequency 
\cite{Mostepanenko-1985, Silva-Farina-PRD-2011}.
Moreover, when considering a time-dependent parameter $\mu(t)$, as given in Eq. 
\eqref{eq:time-dependent-robin-bc}, one has the simulation of a 
perfectly reflecting metal with a time-dependent plasma frequency \cite{Silva-Farina-PRD-2011}.
Since Eq. \eqref{eq:time-dependent-robin-bc} is a
particular case of our model given in Eq. \eqref{eq:model},
the Lagrangian proposed here can simulate a partially reflecting metal with a time-dependent plasma frequency. 

In the next section, we compute and discuss the spectrum and the total
number of created particles from Eqs. (\ref{eq:out-in}), (\ref{eq:Scattering-matrix})
and (\ref{eq:S-full}).

\section{\label{sec:spectrum}Particle Creation}

Considering vacuum as the initial state of the field, the spectrum
of created particles can be computed by \cite{Lambrecht-Jaekel-Reynaud-PRL-1996}
\begin{equation}
N(\omega)=2\omega\,\mathrm{Tr}\left\langle 0_{\text{in}}\right|\Phi_{\text{out}}(-\omega)\Phi_{\text{in}}^{\mathrm{T}}(\omega)\left|0_{\text{in}}\right\rangle .
\label{eq:N-def}
\end{equation}
Substituting Eq. (\ref{eq:out-in}) into (\ref{eq:N-def}), we obtain
\begin{equation}
N(\omega)=N_{+}(\omega)+N_{-}(\omega),
\label{eq:NmNm}
\end{equation}
where
\begin{equation}
N_{\pm}(\omega)=\frac{\epsilon^{2}\mu_{0}^{2}}{\pi}\int_{0}^{\infty}\frac{\mathrm{d}\omega^{\prime}}{2\pi}\frac{\omega}{\omega^{\prime}}\frac{\mathrm{Re}\left[1+r_{\pm}(-\omega^{\prime})\right]|\tilde{f}(\omega+\omega^{\prime})|^{2}}{\left(\mu_{0}-\chi_{0}\omega^{2}\right)^{2}+\omega^{2}(1+\lambda_{0}^{2})^{2}},
\label{eq:N-pm-0}
\end{equation}
with $N_{+}(\omega)$ and $N_{-}(\omega)$ being the spectra for the
right and left sides of the mirror, respectively.
Manipulating this formula, we obtain
\begin{eqnarray}
N_{\pm}(\omega) & = & \frac{\epsilon^{2}}{\pi}(1\pm\lambda_{0})^{2}(1+\lambda_{0}^{2})\nonumber \\
&  & \times\int_{0}^{\infty}\frac{\mathrm{d}\omega^{\prime}}{2\pi}\Upsilon(\omega)\Upsilon(\omega^{\prime})|\tilde{f}(\omega+\omega^{\prime})|^{2},
\label{eq:Npm}
\end{eqnarray}
where
\begin{equation}
\Upsilon(\omega)=\frac{\mu_{0}\omega}{(\mu_{0}-\chi_{0}\omega^{2})^{2}+\omega^{2}(1+\lambda_{0}^{2})^{2}}.
\end{equation}
Considering $\mu_{0}\rightarrow\infty$ or $\chi_{0}\rightarrow\infty$,
we have $N_{\pm}(\omega)\rightarrow0$, a Dirichlet
BC on both sides of the mirror, as mentioned in the previous section
[Eq. (\ref{eq:Dirichlet-L-R})].
We also remark the symmetry $\lambda_0\leftrightarrow-\lambda_0$ for 
$N(\omega)$,
\begin{equation}
	N(\omega)\big|_{\lambda_0}=N(\omega)\big|_{-\lambda_0}.
	\label{eq:N-lambda-0-symmetry}
\end{equation}

From Eq. (\ref{eq:Npm}), we conclude that
\begin{equation}
N_{-}(\omega)=\left(\frac{1-\lambda_{0}}{1+\lambda_{0}}\right)^{2}N_{+}(\omega).
\label{eq:N-N}
\end{equation}
Therefore, the spectra, for each side of the mirror, differ from each
other only by a frequency-independent global factor. 
For $\lambda_{0}>0$,
$N_{-}(\omega)$ is smaller than $N_{+}(\omega)$ for all frequencies,
and the opposite occurs for $\lambda_{0}<0$. The spectra are symmetric
i.e., $N_{-}(\omega)=N_{+}(\omega)$ only if $\lambda_{0}=0$,
or in the limit $\lambda_{0}\rightarrow\infty$. Furthermore, from
Eq. (\ref{eq:N-N}), we can conclude that there will not be particle
creation for one of the sides of the mirror if $\lambda_{0}\rightarrow\pm1$,
specifically $N_{-}(\omega)=0$ if $\lambda_{0}\rightarrow1$, 
or $N_{+}(\omega)=0$
if $\lambda_{0}\rightarrow-1$, which is a consequence of the fact
that, in these limits, the field obeys the (time-independent) Dirichlet
BC for one of the sides and the time-dependent generalized Robin BC
on the other side {[}see Eqs. (\ref{eq:Dir}) and (\ref{eq:time-dependent-generalized-Robin-BC}){]}.

The total number of created particles is obtained by integrating
the spectrum for all frequencies,
\begin{equation}
\mathcal{N}=\int_{0}^{\infty}\mathrm{d}\omega N(\omega).
\label{eq:Total-N}
\end{equation}
From Eq. (\ref{eq:NmNm}), the last equation can be written as $\mathcal{N}=\mathcal{N}_{+}+\mathcal{N}_{-}$,
where $\mathcal{N}_{+}$ and $\mathcal{N}_{-}$ are the total number
of particles for the right and left sides of the mirror, respectively.
From Eq. (\ref{eq:N-N}), it follows that
\begin{equation}
\frac{\mathcal{N}_{-}}{\mathcal{N}_{+}}=\left(\frac{1-\lambda_{0}}{1+\lambda_{0}}\right)^{2}.
\label{eq:N-N-Rate}
\end{equation}
From Eq. \eqref{eq:N-lambda-0-symmetry}, we have
\begin{equation}
\mathcal{N}\big|_{\lambda_0}=\mathcal{N}\big|_{-\lambda_0}.
	\label{eq:N-total-lambda-0-symmetry}
\end{equation}

The total energy $\mathcal{E}$, dissipated from the mirror and converted in real particles, is given by
\begin{equation}
\mathcal{E}=\int_{0}^{\infty}\mathrm{d}\omega N(\omega)\omega = \mathcal{E}_{+} + \mathcal{E}_{-},
\label{eq:Total-E}
\end{equation}
where
\begin{equation}
\mathcal{E}_{\pm}=\int_{0}^{\infty}\mathrm{d}\omega N_{\pm}(\omega)\omega.
\label{eq:Total-E-pm}
\end{equation}

The integral in Eq. (\ref{eq:Total-N}) is suitable for numerical
integration and we shall discuss these results in the following.

\section{\label{sec:application} Application}

From now on, we consider, in Eq. \eqref{eq:mu-of-t}, 
the time-varying behavior given by
\begin{equation}
f(t)=\cos(\omega_{0}t)\exp(-|t|/\tau),
\label{eq:f}
\end{equation}
where $\tau$ is the time interval for which the oscillations occur
effectively and $\omega_{0}$ is the oscillation frequency. 
We also
consider $\omega_{0}\tau\gg1$, called monochromatic
limit \cite{Silva-Braga-Rego-Alves-PRD-2015}. 
This is a typical oscillatory behavior considered in investigating
the DCE \cite{Johansson-PRL-2009,Johansson-et-al-PRA-2010, Silva-Farina-PRD-2011, Silva-Braga-Rego-Alves-PRD-2015}.
The Fourier transform of $f(t)$, considering the monochromatic limit,
is given by \cite{Silva-Braga-Rego-Alves-PRD-2015}
\begin{equation}
|\tilde{f}(\omega)|/\tau=(\pi/2)[\delta(\omega+\omega_{0})+\delta(\omega-\omega_{0})].
\label{eq:monochromatic}
\end{equation}
Therefore, substituting Eq. (\ref{eq:monochromatic}) into (\ref{eq:Npm})
we obtain the following expression for the spectra
\begin{equation}
\frac{N_{\pm}(\omega)}{\tau}=
\frac{\epsilon^{2}(1\pm\lambda_{0})^{2}(1+\lambda_{0}^{2})}{4\pi}
\Upsilon(\omega)\Upsilon(\omega_{0}-\omega)\Theta(\omega_{0}-\omega).
\label{eq:Spectra}
\end{equation}

The spectrum in Eq. \eqref{eq:Spectra} presents the symmetry 
\begin{equation}
	N_{\pm}(\omega_{0}/2+\zeta)=N_{\pm}(\omega_{0}/2-\zeta),
	\label{eq:Sym-Spec}
\end{equation}
where $|\zeta|<\omega_{0}/2$. 
Taking this symmetry into account in \eqref{eq:Total-E-pm}, we have 
a proportionality between energy and number of particles, given by
\begin{equation}
	\mathcal{E}_{\pm}=\frac{\omega_{0}}{2}\mathcal{N}_{\pm},
	\label{eq:Total-E-pm-as-factor-of-N}
\end{equation}
which leads to 
\begin{equation}
	\mathcal{E}=\frac{\omega_{0}}{2}\mathcal{N}.
	\label{eq:Total-E-as-factor-of-N}
\end{equation}

\begin{figure}[t]
	\includegraphics[width=0.96\columnwidth]{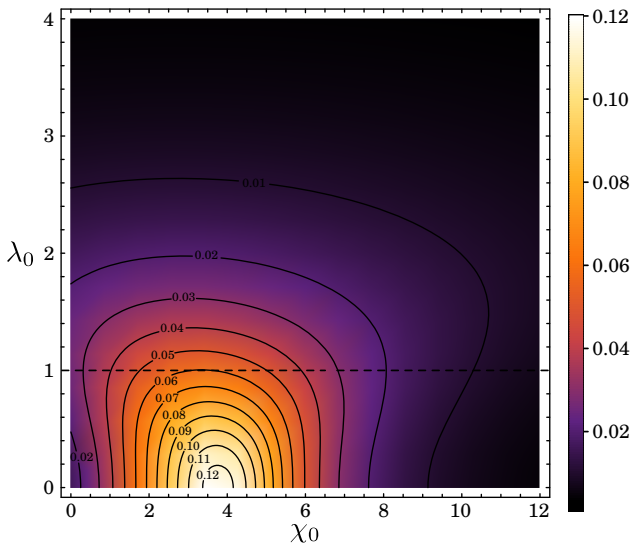}
	\caption{The normalized total number of created particles $(2\pi /\epsilon^2\tau)\mathcal{N}$, as a function of the parameters $\chi_{0}$ and $\lambda_{0}$, considering $\mu_{0}=1$ and $\omega_{0}=1$. The dashed and solid lines represent level curves.
	This figure contains, as particular cases, some results found in the literature.
	The dashed line corresponds to the case $\lambda_{0}=1$,
	and shows $(2\pi /\epsilon^2\tau)\mathcal{N}$ for a perfectly reflecting mirror that imposes a generalized time-dependent Robin BC to the field (result found in Ref. \cite{Rego-Silva-Alves-Farina-PRD-2014}).
	The point $(\chi_{0}=0,\lambda_{0}=1)$ shows $(2\pi /\epsilon^2\tau)\mathcal{N}$ 
	for a perfectly reflecting Robin BC (result found in Ref. \cite{Silva-Farina-PRD-2011}). 
	The vertical line given by $\chi_{0}=0$
	shows $(2\pi /\epsilon^2\tau)\mathcal{N}$ 
	for the model used in Ref. \cite{Silva-Braga-Rego-Alves-PRD-2020}.
}
	\label{Fig.1} 
\end{figure}

In Fig. \ref{Fig.1}, we show the normalized total rate of the number of created particles, $(2\pi /\epsilon^2\tau)\mathcal{N}$, obtained from 
Eqs. \eqref{eq:NmNm}, \eqref{eq:Total-N}, and \eqref{eq:Spectra},
as a function of $\lambda_{0}$ and $\chi_{0}$. 
Before we begin our analysis of the general aspects of
Fig. \ref{Fig.1}, let us highlight that 
it contains, as particular cases, some results found in the literature \cite{Silva-Farina-PRD-2011,Silva-Braga-Alves-PRD-2016}.
%,
Specifically, the point $(\chi_{0}=0,\lambda_{0}=1)$ corresponds to the result for the normalized particle creation rate 
for a perfectly reflecting Robin BC, found in Ref. \cite{Silva-Farina-PRD-2011}. 
The dashed line in Fig. \ref{Fig.1} corresponds to the case $\lambda_{0}=1$,
and indicates the result for the normalized particle creation rate for a perfectly reflecting mirror that imposes a generalized time-dependent Robin BC to the field, found in Ref. \cite{Rego-Silva-Alves-Farina-PRD-2014}.
Moreover, the vertical line given by $\chi_{0}=0$
corresponds to the normalized particle creation rate for
the model used in Ref. \cite{Silva-Braga-Rego-Alves-PRD-2020},
when the oscillatory behavior given in Eq. \eqref{eq:f} is considered.

In Fig. \ref{Fig.1}, a point in a lighter region has a greater 
normalized particle creation than a point in a darker one.
The dashed line $(\lambda_{0}=1)$ indicates the 
particle creation for perfectly reflecting mirrors,
whereas any point not belonging to this line represents the particle creation for 
partially reflecting ones.
One can find outside the dashed line points lighter than some belonging to it.
This means that partially reflecting static mirrors, with time-varying properties, can produce a larger number of particles than perfectly reflecting ones.
In fact, one can observe peak of particle creation around $(\chi_{0}\approx 3.5,\lambda_0=0)$.
Moreover, using Eq. \eqref{eq:Total-E-as-factor-of-N},
from Fig. \ref{Fig.1} one can have a direct visualization 
of the behavior of $(2\pi /\epsilon^2\tau)\mathcal{E}$,
the normalized rate of total  energy.

\begin{figure}[t]
	\includegraphics[width=0.96\columnwidth]{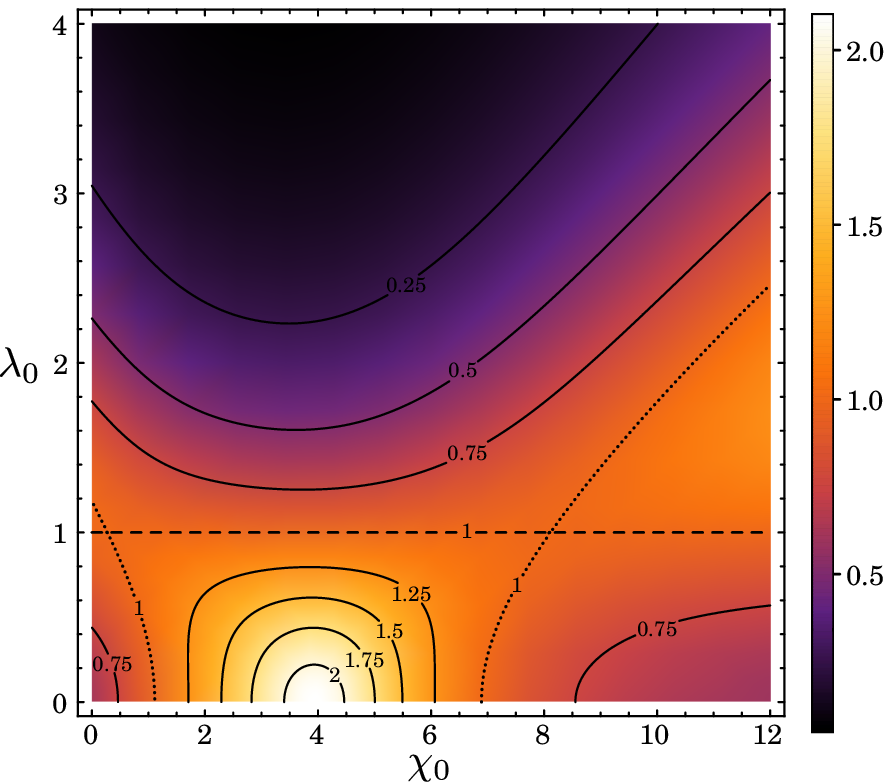}
	\caption{The ratio $\mathcal{N}/\mathcal{N}|_{\lambda_{0}=1}$ as a function of the parameters $\chi_{0}$ and $\lambda_{0}$, considering $\mu_{0}=1$ and $\omega_{0}=1$. The dashed, dotted and solid lines represent level curves.
	This figure contains, as particular cases, results of some models found in the literature.
	The dashed line shows $\mathcal{N}/\mathcal{N}|_{\lambda_{0}=1}=1$, corresponding to the case of a perfectly reflecting mirror that imposes a generalized time-dependent Robin BC to the field (model found in Ref. \cite{Rego-Silva-Alves-Farina-PRD-2014}).
	The point $(\chi_{0}=0,\lambda_{0}=1)$ corresponds to a perfectly reflecting Robin BC (model found in Ref. \cite{Silva-Farina-PRD-2011}). 
	The vertical line given by $\chi_{0}=0$ shows $\mathcal{N}/\mathcal{N}|_{\lambda_{0}=1}$
	for the model used in Ref. \cite{Silva-Braga-Rego-Alves-PRD-2020}.	
}
	\label{Fig.2} 
\end{figure}

For a better visualization of where, in the configuration space $\chi_0\times\lambda_0$, 
a partially reflecting static mirror produces a larger number of particles than the
correspondent perfectly reflecting one, 
in Fig. \ref{Fig.2} we show the ratio $\mathcal{N}/\mathcal{N}|_{\lambda_{0}=1}$.
Moreover, using Eq. \eqref{eq:Total-E-as-factor-of-N},
we can write $\mathcal{N}/\mathcal{N}|_{\lambda_{0}=1}=\mathcal{E}/\mathcal{E}|_{\lambda_{0}=1}$,
so that Fig. \ref{Fig.2} also shows the ratio for the total energy.
Level curves with values $\mathcal{N}/\mathcal{N}|_{\lambda_{0}=1}>1$ indicate
that partially reflecting static mirrors, with a time-varying parameter $\mu(t)$, can produce a larger number of particles than the correspondent perfectly reflecting one.
One can observe greater values of $\mathcal{N}/\mathcal{N}|_{\lambda_{0}=1}$, for instance, around $(\chi_{0}\approx 4,\lambda_0=0)$. 
The dashed line (level curve $=1$) indicates perfectly reflecting mirrors. We highlight that
the dotted lines show a family of semi-transparent mirrors generating a same number of particles
than perfect mirrors.

In order to interpret these results, we remark that $N_{\pm}(\omega)$, in Eq. \eqref{eq:N-pm-0}, 
depends on the reflectivity $|r_{\pm}(\omega^{\prime})|$ and also on
the phase $\text{arg}[r_{\pm}(\omega^{\prime})]$ [see Eq. \eqref{r-pm}].
Mirrors with $|r_{\pm}(\omega^{\prime})|=1$
(ideal mirrors, correspondent to the dashed lines in Figs. \ref{Fig.1} and \ref{Fig.2}) have
a maximum reflectivity, but not necessarily the
combination of $|r_{\pm}(\omega^{\prime})|=1$ and $\text{arg}[r_{\pm}(\omega^{\prime})]$ 
that creates a maximum of particles.
On the other hand, we show that there are points not belonging to 
the dashed lines for which the combination of
$|r_{\pm}(\omega^{\prime})|\neq 1$ and $\text{arg}[r_{\pm}(\omega^{\prime})]$ 
can produce the same or even a greater number of particles than perfectly reflecting mirrors.

%%%%%%%%%%%%%%%%%%%%%%%%%%%%%%%%%%%%%%%%%%%%%%%%%%%%%%%%%%%%%%%%%%
\section{\label{sec:final-remarks} Final Remarks}

The creation of real particles by excitation of the quantum vacuum can be caused,
for instance, by the movement of a mirror \cite{Moore-1970}, or by the time-dependent properties of a static material medium \cite{Yablonovitch-PRL-1989, Lozovik-PZhETF-1995}.
In the context of moving mirrors, 
it had already been shown that transparency can enhance the number of created particles from vacuum, when compared to a perfect mirror \cite{Silva-Braga-Alves-PRD-2016}.
In the present paper, we showed that, even in the context of static mirrors
with time-dependent properties, a partially reflecting mirror can produce
a greater number of particles than a perfectly reflecting one.

%%%%%%%%%%%%%%%%%%%%%%%%%%%%%%%%%%%%%%%%%%%%%%%%%%%%%%%

\begin{acknowledgments}
The authors thank Danilo Pedrelli for valuable discussions, 
and also Lucas Queiroz for 
the support in improving the figures presented here.
\end{acknowledgments}

\end{document}